\documentclass[conference]{IEEEtran}

\usepackage{cite}
\usepackage{amsmath,amssymb,amsfonts}
\usepackage{algorithmic}
\usepackage{graphicx}
\usepackage{subcaption}
\usepackage{textcomp}
\usepackage{xcolor}
\usepackage{booktabs}
\usepackage{multirow}
\usepackage{pifont}
\usepackage{tcolorbox}
\usepackage{colortbl}
\usepackage{makecell}
\usepackage{amsfonts}
\usepackage{url}

\def\BibTeX{{\rm B\kern-.05em{\sc i\kern-.025em b}\kern-.08em
    T\kern-.1667em\lower.7ex\hbox{E}\kern-.125emX}}
\begin{document}

\title{Exposing and Defending Membership Leakage in Vulnerability Prediction Models}

\author{
    \IEEEauthorblockN{
        Yihan Liao\IEEEauthorrefmark{1},
        Jacky Keung\IEEEauthorrefmark{1},
        Xiaoxue Ma\IEEEauthorrefmark{2},
        Jingyu Zhang\IEEEauthorrefmark{1}, and
        Yicheng Sun\IEEEauthorrefmark{1}
    }
    \IEEEauthorblockA{
        \IEEEauthorrefmark{1}Department of Computer Science, City University of Hong Kong, Hong Kong, China\\
        \IEEEauthorrefmark{2}Electronic Engineering and Computer Science, Hong Kong Metropolitan University, Hong Kong, China\\
        \{yihanliao4-c, jzhang2297-c, yicsun2-c\}@my.cityu.edu.hk, jacky.keung@cityu.edu.hk, kxma@hkmu.edu.hk
    }
}

\maketitle

\renewcommand{\thefootnote}{\fnsymbol{footnote}}
\footnotetext[2]{Xiaoxue Ma is the corresponding author.}
\renewcommand{\thefootnote}{\arabic{footnote}}

\begin{abstract}
Neural models for vulnerability prediction (VP) have achieved impressive performance by learning from large-scale code repositories. However, their susceptibility to Membership Inference Attacks (MIAs), where adversaries aim to infer whether a particular code sample was used during training, poses serious privacy concerns. While MIA has been widely investigated in NLP and vision domains, its effects on security-critical code analysis tasks remain underexplored. In this work, we conduct the first comprehensive analysis of MIA on VP models, evaluating the attack success across various architectures (LSTM, BiGRU, and CodeBERT) and feature combinations, including embeddings, logits, loss, and confidence. Our threat model aligns with black-box and gray-box settings where prediction outputs are observable, allowing adversaries to infer membership by analyzing output discrepancies between training and non-training samples. The empirical findings reveal that logits and loss are the most informative and vulnerable outputs for membership leakage. Motivated by these observations, we propose a Noise-based Membership Inference Defense (NMID), which is a lightweight defense module that applies output masking and Gaussian noise injection to disrupt adversarial inference. Extensive experiments demonstrate that NMID significantly reduces MIA effectiveness, lowering the attack AUC from nearly 1.0 to below 0.65, while preserving the predictive utility of VP models. Our study highlights critical privacy risks in code analysis and offers actionable defense strategies for securing AI-powered software systems.
\end{abstract}


\section{Introduction} \label{sec:introduction}
Machine learning–based Vulnerability Prediction (VP) has become a promising direction for assisting static analysis in secure software development. By learning patterns of vulnerable code from annotated datasets, VP models have achieved considerable success across diverse neural architectures, including recurrent models like Long Short-Term Memory (LSTM) and Gated Recurrent Unit (GRU) \cite{zhou2019devign, li2021sysevr}, and pre-trained transformers like CodeBERT \cite{fu2022linevul}. These models are capable of automatically predicting whether a given function or code snippet contains security flaws, and are increasingly integrated into modern development and security \cite{lin2020software}. 

While classification-based VP models have become increasingly prevalent in secure software development workflows, they are often trained on proprietary or sensitive codebases, including private repositories or curated industry datasets, raising privacy concerns when deployed via public APIs or shared checkpoints \cite{chen2023diversevul}. Prior studies have shown that machine learning models can inadvertently memorize and leak training-specific signals, especially when outputs such as logits or losses are externally accessible \cite{carlini2021extracting, feldman2020does}. In software engineering, recent work has also explored how code-based models may serve as unintended leakage channels through public APIs \cite{tramer2022truth}. A particularly concerning yet understudied threat in this context is the Membership Inference Attack (MIA), where an adversary attempts to determine whether a specific code sample was part of a model’s training data \cite{shokri2017membership, hu2022membership}. Although MIAs have been widely studied in domains like vision \cite{liao2025evaluate} and natural language processing (NLP) \cite{duan2024membership}, their application to code-based models remains nascent. Recent efforts have focused on generative code models such as code completion \cite{yang2024gotcha, wan2024does}, which emit token sequences and require sequence-level reasoning. In contrast, we study classification-based VP models, which output discrete labels or class probabilities, signals that are more accessible and potentially more informative for adversaries. Furthermore, the discrete syntax and sparse supervision of code introduce unique membership leakage patterns not commonly seen in vision or NLP models \cite{casey2025survey}.

In this paper, we present the first empirical study of assessing the vulnerability of VP models to MIAs. We conduct comprehensive evaluations on two datasets, including Software Assurance Reference Dataset (SARD) \cite{nist_sard} and National Vulnerability Database (NVD) \cite{nist_nvd}, across multiple VP architectures, including bidirectional LSTM (BiLSTM) \cite{guo2023detecting}, bidirectional GRU (BiGRU) \cite{jeon2021autovas}, and CodeBERT \cite{feng2020codebert}. Using a suite of eight representative feature combinations derived from model outputs (logits, confidence score, loss, and embedding representations), we rigorously evaluate MIA performance under black-box and gray-box settings with Multilayer Perceptron (MLP) and Convolutional Neural Network (CNN)-based attack models \cite{hu2022membership, wu2024rethinking}. Through these experiments, we reveal that VP models are highly susceptible to MIAs, especially when adversaries leverage intermediate outputs such as logits and loss, which consistently yield the highest attack performance. Particularly, when logits and loss are combined with confidence scores, the attack F1-score can exceed 0.95 across three VP model architectures. We observe a clear separation between member and non-member samples in the feature space, indicating that the model encodes membership-related patterns that MIAs can exploit. These findings demonstrate the urgent need for effective defenses and motivate our design of NMID, a lightweight yet robust mechanism for mitigating MIA. 

NMID does not require retraining VP models, instead, it applies lightweight masking and smoothing mechanisms to logits and loss vectors at inference time, adding randomized noise or suppressing dominant signals that the adversary may exploit. We conduct ablation studies to examine the effectiveness of each component in NMID. Results show that simple masking alone offers limited protection, but the introduction of smoothing, implemented as additive Gaussian noise, substantially reduces attack success. When applying NMID with a high noise scale ($\alpha=10$), the attack accuracy drops by an average of over 0.35 across both MLP and CNN adversaries. In the strongest defense setting, NMID reduces the attack accuracy from 0.9959 to 0.5781 and the F1-score from 0.9993 to 0.5630 on the CodeBERT model, which is the most MIA-susceptible among our evaluated models, demonstrating near-random performance. These results indicate that NMID effectively disrupts membership signals while preserving model utility, providing a practical defense solution adaptable to various VP models and deployment settings.

\textbf{Contributions.} Our main contributions are as follows:

\begin{itemize}
    \item We conduct the first comprehensive empirical study of MIAs on VP models, evaluating eight representative features across three VP model architectures.

    \item We reveal critical factors influencing MIA success, including output modality and feature separability, and use visualizations to identify membership leakage.

    \item We propose \textbf{NMID}, a lightweight, inference-time defense that masks and perturbs model outputs to reduce membership leakage. NMID achieves significant attack mitigation without requiring model retraining.
\end{itemize}

\section{Background and Threat Model} \label{sec:background}

\subsection{Membership Inference Attack}
MIA is a privacy attack in which an adversary attempts to determine whether a particular sample was part of a machine learning model's training set. Formally, given a trained model $f(\theta)$, a data point $x$, and its label $y$, the adversary constructs an attack model $\mathcal{A}$ that predicts the membership status: $\mathcal{A}(f_{\theta}(x), y) \xrightarrow{} \{0,1\}$, where label 1 denotes \textit{member} (i.e., $x \in$ target training set) and 0 denotes \textit{non-member} (i.e., $x \notin$ target training set). To perform MIA, the adversary observes a set of features derived from the model’s output $f_{\theta}(x)$ for a given input $x$. These features typically include the model’s logits \textbf{z} $\in \mathbb{R}^K$, where $K$ denotes the number of output classes and each $z_i$ represents the unnormalized score assigned to class $i$ \cite{he2025towards}. From these logits, the adversary can compute the confidence, defined as $max_i$ Softmax$(z_i)$, where Softmax$(z_i) \in (0, 1)$ and $\sum_i$ Softmax$(z_i) = 1$, reflecting the model’s certainty in its prediction \cite{liu2024please}. Additionally, the loss value $\ell
(f_{\theta}(x), y)$ often computed using cross-entropy between the predicted distribution and the true label $y$, provides information about how well the model fits the input \cite{duan2024membership}. In some settings, intermediate representations such as embeddings $\textbf{e}_x$ obtained from hidden layers (e.g., the [CLS] token output in CodeBERT) are also accessible and serve as rich sources of information about the input \cite{bai2024membership}. By analyzing logits, confidence, loss, and embeddings, the adversary attempts to infer whether the input $x$ was part of the training set, based on the assumption that the model behaves differently on members and non-members.

\subsection{Vulnerability Prediction Models}
VP aims to determine whether a given code snippet contains security flaws. Formally, given a token sequence $x = [t_1, t_2, ..., t_n]$, a model $f_\theta$ predicts a label $y \in {0, 1}$, indicating whether the function is vulnerable. The input is first embedded as $\textbf{E} = [\textbf{e}_1, ..., \textbf{e}n] \in \mathbb{R}^{n \times d}$, using randomly initialized or pre-trained embeddings (e.g., CodeBERT \cite{guo2020graphcodebert}). The embedded sequence is encoded using a neural architecture, such as BiLSTM, BiGRU \cite{jeon2021autovas}, or Transformer, all of which are widely adopted in VP tasks. We include BiLSTM and BiGRU as representative RNN-based models for their strong sequential modeling capabilities, and CodeBERT as a Transformer-based pre-trained model that leverages global attention to capture long-range dependencies and rich contextual semantics. This architectural diversity enables a comprehensive evaluation of MIA vulnerabilities across both lightweight RNNs and large-scale pretrained models. For RNN-based models, the final hidden state $\textbf{h} \in \mathbb{R}^h$ is used to summarize the sequence, while Transformer models utilize the [CLS] token representation. This vector is then passed to a classifier head $g_\phi(h)$ to produce prediction logits $\textbf{z} \in \mathbb{R}^2$, followed by a softmax layer and cross-entropy loss $\mathcal{L}(y, \textbf{z})$ for training. During inference and evaluation, outputs such as logits, confidence scores, and loss are often exposed \cite{wan2024does, yang2024gotcha}. While useful for auditing, they may unintentionally leak membership signals, motivating our investigation of MIA in VP.

\subsection{Threat Model}
We consider a privacy threat where an adversary aims to determine whether a given code snippet was part of the training data of a deployed VP model. Specifically, the target model has been trained on a private dataset and is accessible for inference. The adversary has a set of inputs $x$ with known labels $y$, and aims to infer the membership status of each sample concerning the target training set. We evaluate MIAs under two realistic threat settings:

\textbf{Black-box Setting.} The adversary can only interact with the deployed model through query access and observe per-sample outputs such as logits $\textbf{z}$, confidence score (i.e., $\max_i$ Softmax$(z_i)$), and the cross-entropy loss $\ell$ computed using the true label. This reflects the typical scenario where model predictions and confidence scores are exposed for interpretability or auditing purposes, but internal model parameters and representations  (e.g., model weights, intermediate hidden states) remain inaccessible \cite{liu2024please, duan2024membership}.

\textbf{Gray-box Setting.} In addition to the outputs available in the black-box case, the adversary also gains access to the intermediate feature representations of the input sample, such as hidden layer $\textbf{e}_x$ or the [CLS] token output in transformer-based models. This simulates settings where the model is internally deployed (e.g., in shared organizational environments) or partially exposed via debugging APIs \cite{bai2024membership}.

In both settings, the adversary uses these features to train a binary classifier that distinguishes between member and non-member samples. The feasibility of such attacks is predicated on the assumption that models behave differently on training data (members) and unseen samples (non-members). This behavioral gap, if left unaddressed, can lead to privacy leakage, even in the absence of direct access to model parameters.

\section{Methodology} \label{sec:methodology}
\subsection{Overview of Attack Framework}
\begin{figure}
    \centering
    \includegraphics[width=0.8\linewidth]{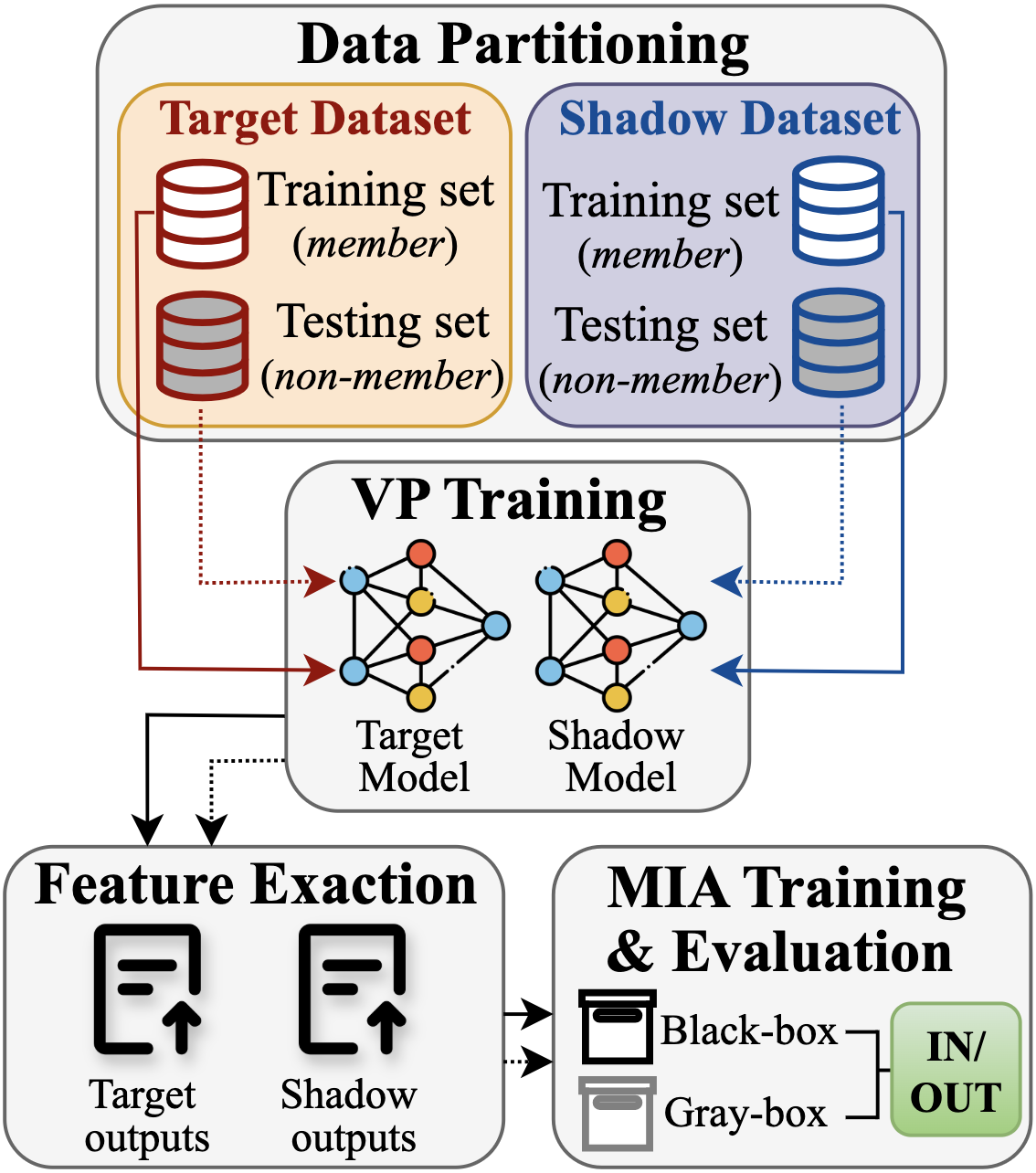}
    \caption{Overview of the Shadow-based MIA Pipeline. Solid arrows indicate the training stage, dashed arrows represent the evaluation stage.}
    \label{fig:attack_framework}
\end{figure}

We adopt a shadow-based MIA pipeline, as illustrated in Figure \ref{fig:attack_framework}. The overall process consists of four key components across both training and evaluation stages: \textit{\textbf{(1) Data Partitioning:}} The input dataset is randomly divided into two disjoint subsets, the target and shadow datasets, each of which is further split into training (member) and testing (non-member) sets. \textit{\textbf{(2) VP Training:}} The target model is trained on the target training set, while a shadow model is trained on the shadow training set to mimic the target model’s behavior. \textit{\textbf{(3) Feature Extraction:}} Both trained models generate outputs (e.g., logits, loss, confidence) for their respective member and non-member samples, forming the feature space for attack modeling. \textit{\textbf{(4) MIA Training and Evaluation:}} Based on the shadow model's outputs, an adversary trains an attack model to distinguish members from non-members. This attack model is then evaluated against the target model's outputs in the black-box or gray-box setting.

This shadow training strategy aligns with prior MIA methodologies \cite{shokri2017membership} and supports evaluation under both black-box and gray-box settings, depending on the extent of model outputs accessible to the adversary. In the black-box case, only externally observable outputs (e.g., loss, logits) are utilized, whereas in gray-box settings, access to intermediate representations (e.g., embeddings) may also be allowed. To simulate a realistic deployment scenario, the attack model is trained on the outputs of a locally trained shadow model and evaluated on the outputs of the unseen target model. The attack is considered successful if it can correctly infer whether a given sample was part of the target model's training data.

\subsection{Feature Construction}
To systematically explore how different types and quantities of model-derived information influence MIA success, we construct eight feature combinations, denoted as \textit{Feature 1} through \textit{Feature 8}, each capturing varying degrees of accessibility and informativeness. These features are summarized in Table \ref{tab:feature_combinations}, and span from strictly black-box signals, such as confidence, loss values, and raw logits, to gray-box settings that include hidden representations (e.g., token-level embeddings). We denote $K$ as the number of output classes and $d$ as the size of the embedding vector.

\begin{table}[h]
\centering
\caption{Feature combinations used for MIA.}
\label{tab:feature_combinations}
\resizebox{0.48\textwidth}{!}{
\begin{tabular}{lllll}
\toprule
\textbf{Feature ID} & \textbf{Combination} & \textbf{Access Type} & \textbf{Dimension} & \textbf{Description} \\
\midrule
Feature 1 & Confidence only                   & Black-box & 1           & Max softmax probability \\
Feature 2 & Loss only                         & Black-box & 1           & Cross-entropy loss      \\
Feature 3 & Logits only                       & Black-box & $K$         & Raw class scores        \\
Feature 4 & Logits + Confidence               & Black-box & $K+1$       & Class-wise + max prob   \\
Feature 5 & Logits + Loss                     & Black-box & $K+1$       & Class-wise + loss       \\
Feature 6 & Logits + Confidence + Loss        & Black-box & $K+2$       & Richer decision signals \\
Feature 7 & Embedding + Logits                & Gray-box  & $d + K$     & Intermediate + output   \\
Feature 8 & Embedding + Logits + Loss         & Gray-box  & $d + K + 1$ & Most informative combo  \\
\bottomrule
\end{tabular}}
\end{table}

In black-box scenarios (\textit{Feature 1} – \textit{Feature 6}), the adversary is assumed to have access only to standard model outputs returned during inference. These settings reflect realistic deployment cases, such as models served via APIs or integrated into IDE-based plugins. For example, \textit{Feature 4} combines logits with the confidence, offering insight into both class-level score distributions and the model's overall prediction certainty. In gray-box settings (\textit{Feature 7} and \textit{Feature 8}), the adversary additionally obtains intermediate-layer embeddings that precede the final classification layer. For CodeBERT, we follow common practice and extract the [CLS] token embedding as the fixed-length sample representation, which is also directly used for classification. For BiLSTM and BiGRU models, we extract the hidden state at the final time step of the recurrent layer, which likewise serves as the input to the classifier. This design aligns the extracted embeddings with each model’s decision-making representation, eliminating mismatch between classification and MIA features. Using a consistent, task-relevant embedding across architectures allows a fair comparison of leakage patterns and more faithfully reflects the privacy risk inherent in the model’s internal features.

\subsection{Attack Models}
To perform MIA, we employ two widely used neural network structures, namely MLP and CNN, as attack models to distinguish between members and non-members based on the features extracted from the shadow model outputs.

The MLP attack model is a feed-forward neural network that processes input features as flat vectors. It consists of two hidden layers with ReLU activations, followed by a final softmax layer that outputs membership probabilities. This structure is lightweight and effective for modeling feature distributions in tabular data, and has been extensively used in prior MIA literature \cite{shokri2017membership, nasr2019comprehensive}. The CNN attack model treats the input feature vector as a one-dimensional sequence and applies convolutional filters to capture local patterns among adjacent values. This model can better exploit spatial correlations in the feature space, particularly when the inputs include structured vectors such as logits or embeddings. Previous studies have shown that CNNs are competitive in MIA \cite{liu2024please, duan2024membership}.

Both attack models are trained on shadow model outputs using member and non-member labels, and evaluated on target model outputs to assess the effectiveness of MIA. We use cross-entropy as the training loss and report attack accuracy, precision, recall, F1-score, and area under the ROC curve (AUC) as evaluation metrics. By comparing MLP and CNN across different feature combinations, we aim to understand how attack capacity varies with model input and complexity.

\section{Experimental Setup}

\begin{figure}[t]
  \centering
  \includegraphics[width=1\linewidth]{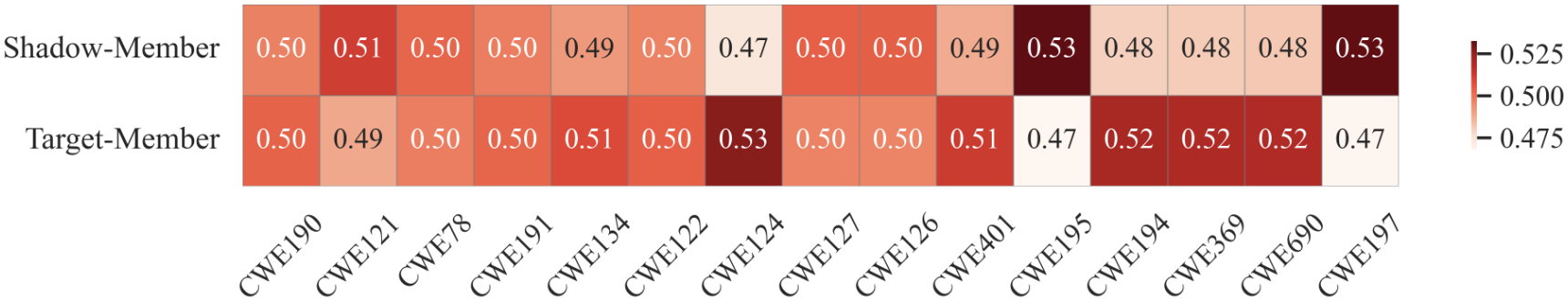}
  \caption{Per-CWE vulnerable label ratios in shadow and target member sets.}
  \label{fig:cwe-distribution}
\end{figure}

\subsection{Dataset}
We construct our experimental dataset by combining two widely used vulnerability datasets: SARD and NVD. SARD provides synthetic C/C++ functions generated by the Juliet Test Suite, each of which is explicitly labeled using the Common Weakness Enumeration (CWE) taxonomy. The NVD dataset comprises real-world vulnerable code snippets, each associated with a CWE identifier obtained from the NVD JSON feed \footnote{\url{https://nvd.nist.gov/developers/vulnerabilities}}, where structured metadata fields provide standardized vulnerability annotations. Aligning NVD’s labeling scheme with that of SARD facilitates consistent categorization and comparison across the two datasets.

After filtering out unlabeled, malformed, or unparsable instances, we obtain 20,000 SARD samples and 2,000 NVD samples. Both datasets contain vulnerable (label 1) and non-vulnerable (label 0) functions, allowing us to construct balanced binary classification subsets. We first partition the combined dataset into non-overlapping shadow and target datasets of equal size. Each dataset is then further divided into member and non-member subsets, also with equal sizes. Within each subset of members and non-members, we ensure a balanced distribution of vulnerable and non-vulnerable samples. This setup avoids label bias and ensures that no raw sample is shared across any of the four subsets to prevent data leakage.

While the shadow and target sets span a similar range of CWE types, the number of samples per CWE is not strictly matched, reflecting realistic distribution shifts across different data sources. To minimize structural drift, we control the aggregate CWE distribution at the corpus level and visualize the CWE-type ratios of vulnerable samples in Figure~\ref{fig:cwe-distribution}. The figure presents the top 15 CWE types, including canonical targets such as CWE-121 (Stack-based Buffer Overflow), CWE-124 (Buffer Underwrite), and CWE-190 (Integer Overflow). The near-identical CWE distributions between the shadow and target datasets support fair MIA evaluation.

\subsection{Vulnerability Prediction Models}
We evaluate three representative neural architectures for VP, including BiLSTM, BiGRU, and CodeBERT. These models encompass a range from simple recurrent encoders to large-scale pre-trained language models, enabling the examination of the generalizability of MIA across various model architectures.

\textbf{BiLSTM.} 
Our LSTM-based model is adapted from common sequence classification architectures \cite{hochreiter1997long}. The input code snippet is tokenized and mapped to a fixed-length vector sequence via an embedding layer. A bidirectional LSTM layer with a hidden dimension of 128 processes the embedded sequence. The final hidden states from both directions are concatenated and passed to a fully connected output layer to produce logits. The model is trained using cross-entropy loss.

\textbf{BiGRU.} 
The BiGRU model follows a similar structure to the LSTM but replaces the recurrent encoder with a GRU \cite{cho2014learning}, which is more parameter-efficient and faster to train. The embedding layer is initialized using a fixed pre-trained embedding matrix. The GRU output at the last time step is passed through a ReLU activation, dropout, and a fully connected classifier layer. To improve generalization, the model integrates optional L1 and L2 regularization terms, which are explicitly included in the training loss. We use this model to extract both sequence embeddings and logits for MIA evaluation \cite{jeon2021autovas}.

\textbf{CodeBERT.} 
CodeBERT is a transformer-based model pre-trained on paired source code and natural language documentation. For classification, the [CLS] token embedding from the final layer is passed through a dropout layer and a dense head to produce logits. Its transformer architecture captures global semantics and long-range dependencies, complementing the local patterns learned by RNN-based models. We follow standard fine-tuning procedures \cite{feng2020codebert} and extract the final-layer token embeddings and logits for constructing MIA features.

\subsection{Training Details}
To support a comprehensive evaluation of MIAs in VP tasks, we trained models using a consistent and reproducible experimental pipeline. An overview of this pipeline is illustrated in Figure \ref{fig:attack_framework} (see Section \ref{sec:methodology}). The training details of our experimental setup are summarized below. 

\textbf{Data Partitioning.} 
We begin by merging the SARD and NVD datasets, resulting in a combined dataset of labeled samples, where each function is annotated as vulnerable or non-vulnerable. The merged data contains 8,757 vulnerable and 13,243 non-vulnerable samples. We then split each category (vul/non-vul) equally into shadow and target sets to ensure disjoint yet distributionally similar partitions, each with a 7:3 train-test ratio. The member samples are used for training the target or shadow models, while both member and non-member test sets are used for evaluating MIA. 

\textbf{VP Model Training.}
To train the VP models on their respective member datasets, we adopt cross-entropy loss and ensure class-balanced training. For token-based models, including BiLSTM and BiGRU, each function is tokenized and encoded as a sequence of integer indices, which are mapped to pre-trained embeddings. The embedding layer remains frozen during training, and the model is optimized using the Adam optimizer with a learning rate of 0.001 and a batch size of 64. We train for up to 100 epochs with early stopping based on validation loss. For CodeBERT, we use the \texttt{codebert-base}\footnote{\url{https://huggingface.co/microsoft/codebert-base}} as the encoder. Input functions are tokenized with a maximum sequence length of 256. CodeBERT models are fine-tuned for 10 epochs with a batch size of 16 using the AdamW optimizer and a learning rate of $2\times10^{-5}$. In both cases, we monitor model performance using accuracy and F1-score.

\textbf{Feature Extraction.}
To simulate MIA, we extract a consistent set of per-sample outputs from the trained VP models, including logits, confidence, cross-entropy loss, and intermediate embeddings. For RNN-based models (BiLSTM, BiGRU), we use the averaged token embeddings as the sample-level representation. For CodeBERT, we extract the [CLS] token embedding from the final hidden layer. These features capture different levels of model behavior, from prediction confidence to internal representations. All features are saved for training downstream attack models, and are grouped into eight combinations (\textit{Feature 1–8}) as summarized in Table~\ref{tab:feature_combinations}.

\textbf{MIA Training.}
After extracting features, we implement two neural architectures for the attack model: MLP and CNN. The MLP consists of two hidden layers (128 and 64 units, respectively) with ReLU activations, followed by a final softmax output layer. The CNN treats the input feature vector as a single-channel temporal signal, applies two convolution layers followed by adaptive max pooling, and outputs predictions through a fully connected classifier. Both models take as input various feature combinations extracted from the target and shadow VP models. During training, input features and corresponding member and non-member labels are loaded and fed into the attack model in batches. Models are trained using cross-entropy loss and optimized with Adam (learning rate 0.0001) for up to 100 epochs. We evaluate the trained attack models on the target dataset features.

\subsection{Evaluation Metrics}
To evaluate the effectiveness of MIA across different attack models and feature combinations, we adopt five widely used classification metrics: accuracy, precision, recall, F1-score, and AUC. These metrics comprehensively assess the attack model’s ability to distinguish member from non-member samples. The $TP$, $FP$, $TN$, and $FN$ are denoted as the number of true positives (members correctly identified), false positives (non-members misclassified as members), true negatives, and false negatives, respectively. The metrics are computed as follows: $\text{Accuracy} = \frac{TP + TN}{TP + TN + FP + FN}$, $\text{Precision} = \frac{TP}{TP + FP}$, $\text{Recall} = \frac{TP}{TP + FN}$, $\text{F1-Score} = 2 \cdot \frac{\text{Precision} \cdot \text{Recall}}{\text{Precision} + \text{Recall}}$. We also report AUC, which reflects the trade-off between $TP$ and $FP$ rates over varying thresholds. A higher AUC indicates stronger discriminative power. All metrics are reported per attack setting to capture both effectiveness and potential privacy risks.

\section{Attack Results and Analysis}
To assess the vulnerability of VP models to MIAs, we conduct a series of empirical evaluations. This section is structured around three research questions:

\begin{itemize}
    \item \textbf{\textit{RQ1:}} Can the shadow model effectively mimic the target model for MIA training?
    \item \textbf{\textit{RQ2:}} Do the outputs of the target model encode distinguishable patterns between member and non-member samples?
    \item \textbf{\textit{RQ3:}} Which output features contribute most to successful MIA, and why?
\end{itemize}

We present the results and insights for each RQ in the following subsections.

\subsection{Shadow Model Performance}
\begin{figure}[t]
\centering
\includegraphics[width=0.8\linewidth]{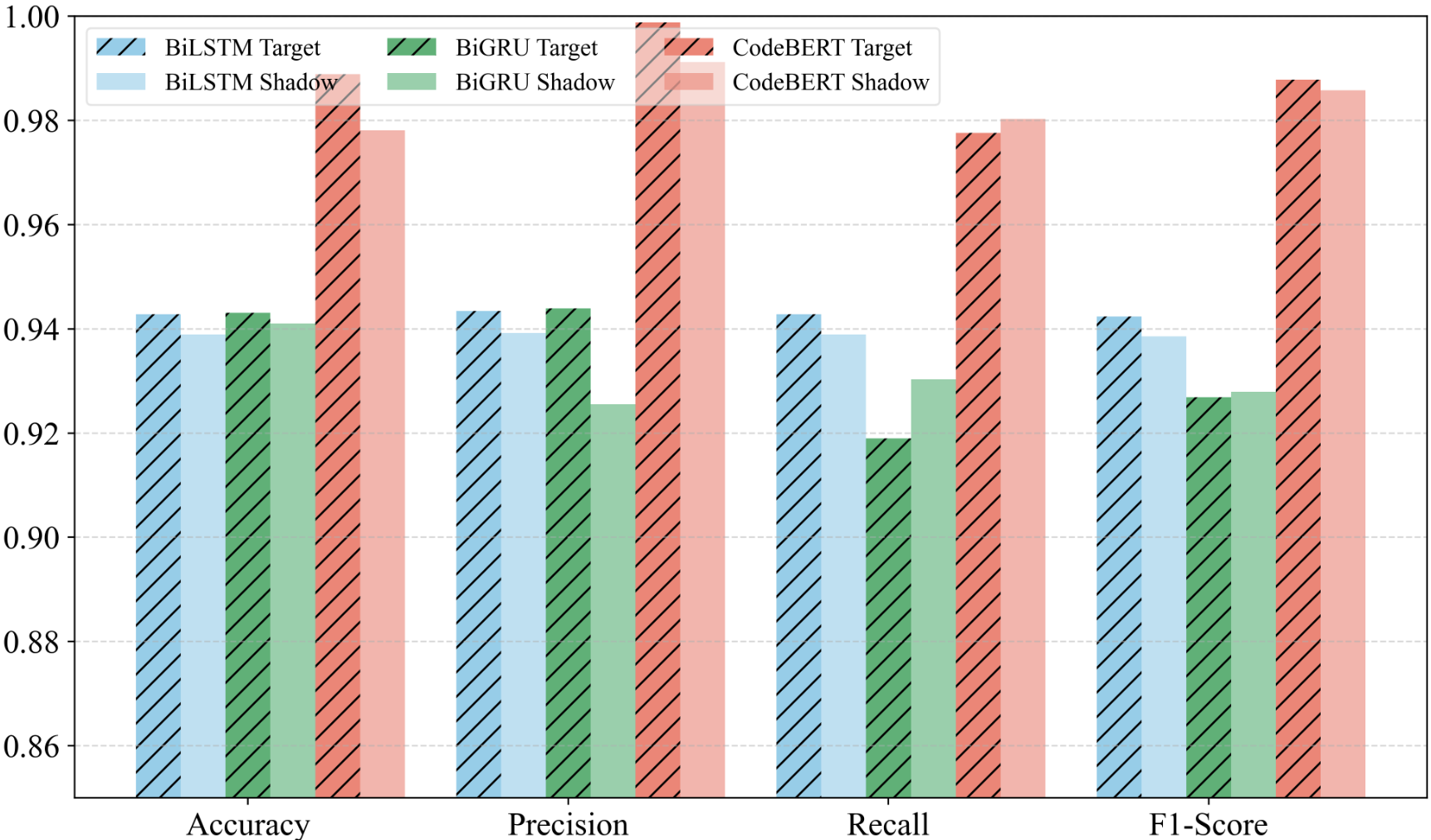}
\caption{Performance of target and shadow VP models across Accuracy, Precision, Recall, and F1-score.}
\label{fig:vp-performance}
\end{figure}

Before performing MIA, we first evaluate the prediction performance of both target and shadow models across three VP models, including BiLSTM, BiGRU, and CodeBERT. As illustrated in Figure \ref{fig:vp-performance}, the shadow models achieve comparable performance among the four metrics to their corresponding target models, indicating that the shadow model successfully mimics the decision boundary of the target model, which is a prerequisite for further MIA training. Specifically, across the three architectures, the differences between shadow and target performance remain consistently small. For BiLSTM and BiGRU, all metric differences are below 0.02. For CodeBERT, the most significant gap is 0.0108 in accuracy. The performance gap between the shadow and target models is negligible, confirming that the shadow models successfully simulate the behavior of their respective target models. Since MIA relies on the assumption that shadow models approximate the behavior of the target model, these results validate the feasibility of using shadow models for downstream MIA training. In addition, all models demonstrate high classification accuracy and F1-scores, especially CodeBERT on both target and shadow sets, likely due to its strong pretraining and capacity to encode rich semantic information. 

\begin{tcolorbox}[colback=gray!30, colframe=gray!70]
    \textbf{\textit{Answer to RQ1}}: Shadow models can effectively approximate the behavior of target models across different VP architectures. It provides a reliable foundation for training attack models in the shadow-based MIA.
\end{tcolorbox}

\begin{table*}[t]
\centering
\caption{Membership Inference Attack Performance using MLP across Different Output Features.}
\label{tab:mlp_mia}
\resizebox{\textwidth}{!}{
\begin{tabular}{c|ccccc|ccccc|ccccc}
\toprule
\multirow{2}{*}{Feature} & \multicolumn{5}{c|}{BiLSTM} & \multicolumn{5}{c|}{BiGRU} & \multicolumn{5}{c}{CodeBERT} \\
& Acc & Pre & Rec & F1 & AUC & Acc & Pre & Rec & F1 & AUC & Acc & Pre & Rec & F1 & AUC \\
\midrule
Feature 1 & 0.5027 & 0.5149 & 0.0944 & 0.1596 & 0.5206 & 0.5000 & 0.5000 & \textbf{1.0000} & 0.6667 & 0.4963 & 0.5012 & \textbf{1.0000} & 0.0024 & 0.0048 & 0.5010 \\
Feature 2 & 0.6093 & 0.6182 & 0.5714 & 0.5939 & 0.6342 & 0.6114 & 0.6170 & 0.5872 & 0.6017 & 0.4688 & 0.6074 & 0.6109 & 0.5920 & 0.6013 & 0.7259 \\
Feature 3 & 0.4979 & 0.4907 & 0.1120 & 0.1824 & 0.5124 & 0.4942 & 0.4967 & 0.8608 & 0.6299 & 0.4880 & 0.5157 & 0.5568 & 0.1544 & 0.2417 & 0.5020 \\
Feature 4 & 0.5000 & 0.5000 & \textbf{0.9988} & 0.6664 & 0.5149 & 0.5064 & 0.5103 & 0.3136 & 0.3885 & 0.5163 & 0.5121 & 0.5478 & 0.1386 & 0.2213 & 0.5092 \\
Feature 5 & 0.9679 & 0.9397 & 0.9379 & 0.9689 & \textbf{0.9999} & 0.9906 & 0.9816 & 0.9980 & 0.9907 & 0.9888 & 0.9894 & 0.9912 & 0.9876 & 0.9894 & \textbf{0.9971} \\
Feature 6 & \textbf{0.9782} & 0.9582 & 0.9979 & \textbf{0.9787} & 0.9996 & \textbf{0.9942} & \textbf{0.9904} & 0.9982 & \textbf{0.9943} & \textbf{0.9999} & 0.9847 & 0.9891 & 0.9806 & 0.9848 & 0.9953 \\
Feature 7 & 0.4836 & 0.4860 & 0.5654 & 0.5227 & 0.4803 & 0.5163 & 0.5257 & 0.3341 & 0.4086 & 0.5116 & 0.4849 & 0.4844 & 0.4703 & 0.4773 & 0.4898 \\
Feature 8 & 0.9643 & \textbf{0.9776} & 0.9504 & 0.9638 & 0.9970 & 0.9682 & 0.9402 & 0.9991 & 0.9692 & 0.9990 & \textbf{0.9985} & 0.9994 & \textbf{0.9976} & \textbf{0.9985} & 0.9927 \\
\bottomrule
\end{tabular}
}
\end{table*}

\begin{table*}[ht]
\centering
\caption{Membership Inference Attack Performance using CNN across Different Output Features.}
\label{tab:cnn_mia}
\resizebox{\textwidth}{!}{
\begin{tabular}{c|ccccc|ccccc|ccccc}
\toprule
\multirow{2}{*}{Feature} & \multicolumn{5}{c|}{LSTM} & \multicolumn{5}{c|}{BiGRU} & \multicolumn{5}{c}{CodeBERT} \\
& Acc & Pre & Rec & F1 & AUC & Acc & Pre & Rec & F1 & AUC & Acc & Pre & Rec & F1 & AUC \\
\midrule
Feature 1 & 0.5000 & 0.0000 & 0.0000 & 0.0000 & 0.4793 & 0.5000 & 0.0000 & 0.0000 & 0.0000 & 0.4991 & 0.5000 & 0.0000 & 0.0000 & 0.0000 & 0.4921 \\
Feature 2 & 0.6071 & 0.6209 & 0.5502 & 0.5834 & 0.7778 & 0.6068 & 0.6157 & 0.5684 & 0.5911 & 0.4680 & 0.6074 & 0.6109 & 0.5920 & 0.6013 & 0.7159 \\
Feature 3 & 0.4893 & 0.4838 & 0.5708 & 0.5237 & 0.4813 & 0.5124 & 0.5191 & 0.3378 & 0.4092 & 0.5068 & 0.5118 & 0.5162 & 0.3759 & 0.4350 & 0.5050 \\
Feature 4 & 0.4815 & 0.4843 & 0.5702 & 0.5238 & 0.4811 & 0.5124 & 0.5186 & 0.3469 & 0.4157 & 0.5051 & 0.5121 & 0.5150 & 0.3856 & 0.4410 & 0.4990 \\
Feature 5 & 0.9610 & 0.9980 & 0.9219 & 0.9594 & \textbf{0.9996} & \textbf{0.9827} & \textbf{0.9978} & \textbf{0.9655} & \textbf{0.9824} & \textbf{0.9999} & \textbf{0.9997} & 0.9962 & 0.9992 & 0.9979 & \textbf{0.9934} \\
Feature 6 & \textbf{0.9622} & \textbf{0.9987} & \textbf{0.9243} & \textbf{0.9607} & 0.9993 & 0.9824 & 0.9899 & 0.9649 & 0.9821 & \textbf{0.9999} & 0.9959 & 0.9992 & \textbf{0.9994} & \textbf{0.9993} & 0.9897 \\
Feature 7 & 0.4812 & 0.4841 & 0.5726 & 0.5247 & 0.4834 & 0.5127 & 0.5189 & 0.3493 & 0.4175 & 0.5153 & 0.5121 & 0.5183 & 0.3420 & 0.4121 & 0.5060 \\
Feature 8 & 0.9588 & 0.9981 & 0.9177 & 0.9571 & 0.9980 & 0.9785 & 0.9975 & 0.9570 & 0.9780 & 0.9996 & 0.9991 & \textbf{0.9993} & 0.9982 & 0.9991 & 0.9827 \\
\bottomrule
\end{tabular}}
\end{table*}

\subsection{Preliminary Distribution Insights into Model Outputs}
Before evaluating MIA, we first examine whether the target model’s outputs differ statistically between member and non-member samples. This analysis serves two purposes: (1) verifying whether the target model overfits to training data, and (2) assessing whether individual output features leak membership information in isolation.

\begin{figure}[t]
    \centering
    \begin{subfigure}[t]{0.48\textwidth}
        \centering
        \includegraphics[width=\textwidth]{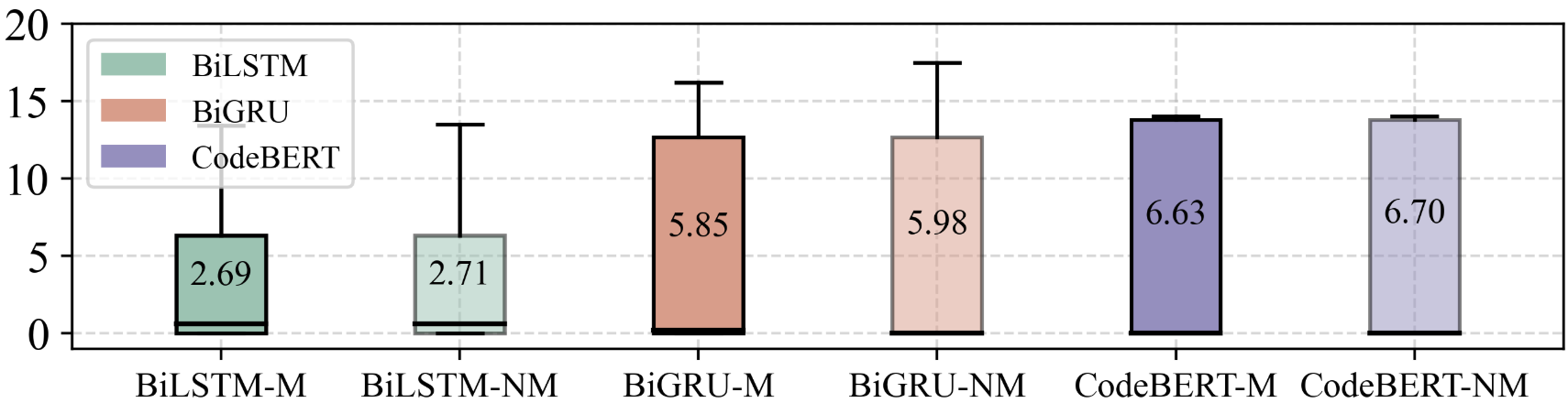}
        \caption{Loss distribution}
        \label{fig:box_loss}
    \end{subfigure}
    \hfill
    \begin{subfigure}[t]{0.48\textwidth}
        \centering
        \includegraphics[width=\textwidth]{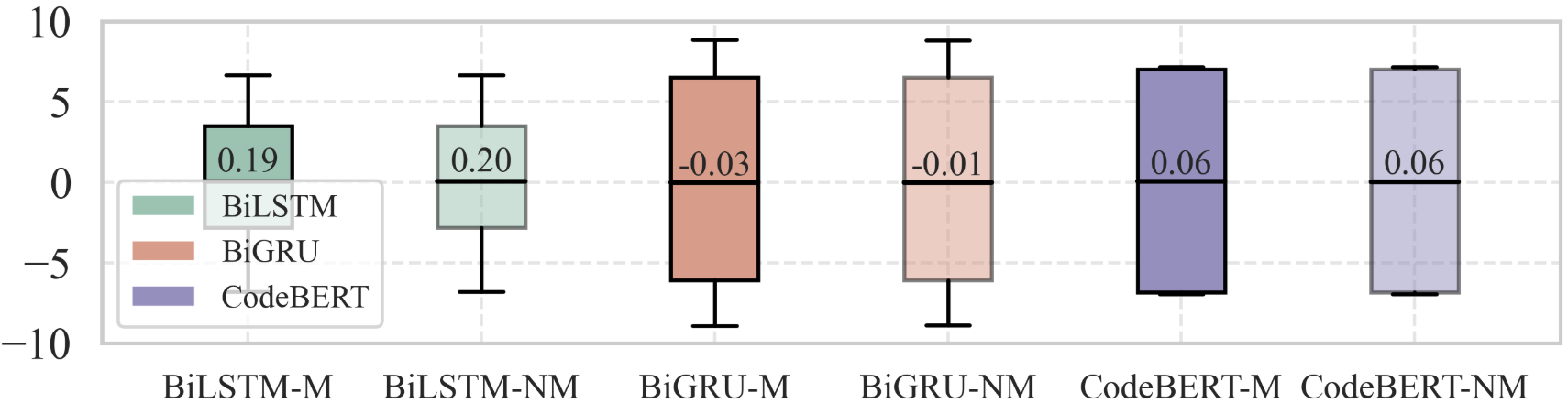}
        \caption{Logits distribution}
        \label{fig:box_logits}
    \end{subfigure}
    \caption{Distribution of loss and logits for member (M) and non-member (NM) samples across BiLSTM, BiGRU, and CodeBERT.}
    \label{fig:box_combined}
\end{figure}

We focus our analysis on loss and logits, as confidence scores are saturated near 1.0 across most samples, offering little discriminative power, and high-dimensional embeddings cannot be meaningfully assessed in one-dimensional plots.
Figure~\ref{fig:box_combined} presents box plots of loss and logits across three VP models: BiGRU, BiLSTM, and CodeBERT. 
For the loss values, we observe a consistent and subtle trend, where member samples exhibit slightly lower median losses than non-members (e.g., 5.85 vs. 5.98 for BiGRU, 6.63 vs. 6.70 for CodeBERT). This aligns with expectations, as training samples are directly optimized during the learning process. However, the differences are small and distributions overlap significantly, indicating that loss alone offers only a weak signal of membership. Logit distributions show even less separation. The median values are nearly identical for members and non-members, with large overlaps in the full distribution. These results suggest that logits are not linearly separable on their own. Therefore, marginal separability indicates that individual output features may not be sufficient for reliable membership inference. Nevertheless, whether these weak signals can be exploited in high-dimensional combinations remains to be evaluated in the following section.

\begin{tcolorbox}[colback=gray!30, colframe=gray!70]
    \textbf{\textit{Answer to RQ2}}: Target model outputs reveal only minor distributional differences between member and non-member samples, suggesting limited information leakage from individual features. 
\end{tcolorbox}

\begin{figure*}[ht]
    \centering
    \begin{subfigure}[t]{\linewidth}
        \centering
        \includegraphics[width=0.9\linewidth]{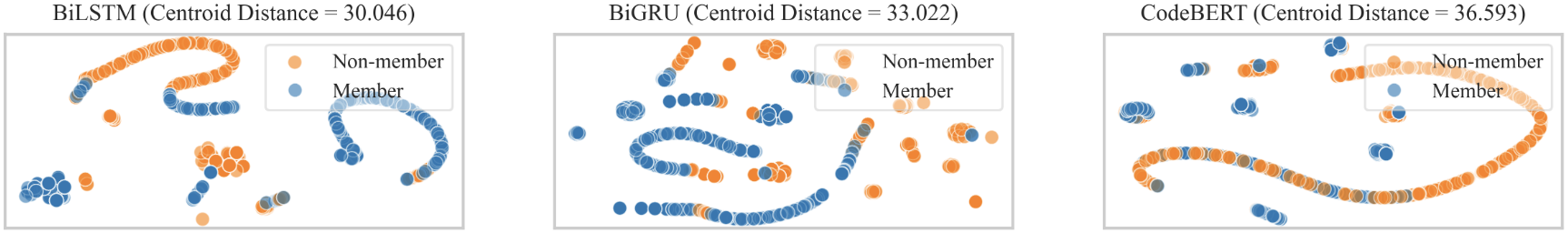}
        \caption{Feature 2 (loss).}
        \label{fig:tsne_loss}
    \end{subfigure}
    
    \begin{subfigure}[t]{\linewidth}
        \centering
        \includegraphics[width=0.9\linewidth]{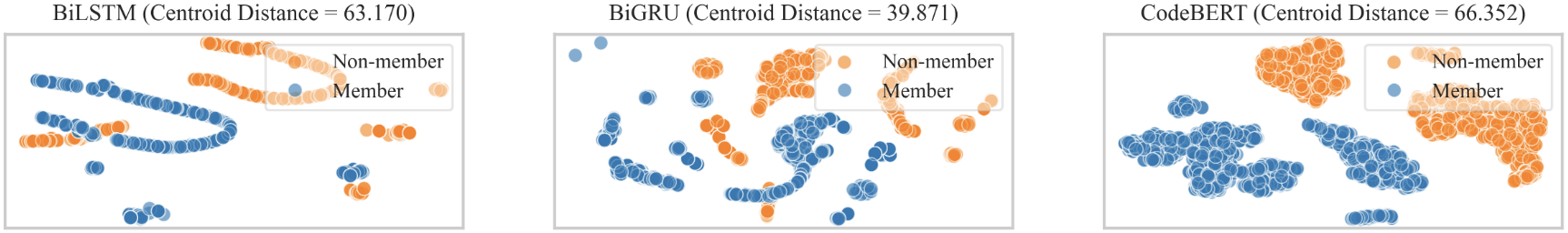}
        \caption{Feature 6 (logits + confidence + loss).}
        \label{fig:tsne_conf}
    \end{subfigure}
    
    \begin{subfigure}[t]{\linewidth}
        \centering
        \includegraphics[width=0.9\linewidth]{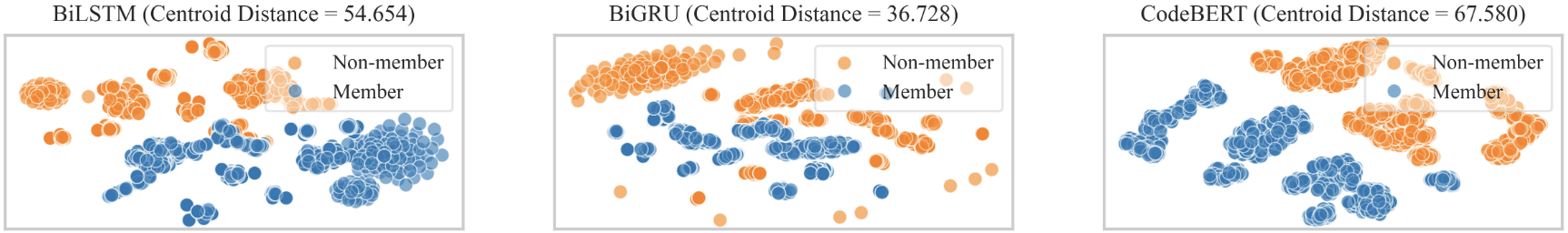}
        \caption{Feature 8 (embedding + logits + loss).}
        \label{fig:tsne_emb}
    \end{subfigure}
    
    \caption{t-SNE visualization of different feature sets (member vs. non-member) on the target testing set. The centroid distance is the average separability between the two classes.}
    \label{fig:tsne-visualization}
\end{figure*}

\subsection{MIA Performance}
Building on the previous analysis, we assess the performance of MIA in this section. We train two types of attack models using various combinations of output features (refer to Table~\ref{tab:feature_combinations}) extracted from the shadow models, and evaluate them on the target model outputs. Tables \ref{tab:mlp_mia} and \ref{tab:cnn_mia} report the results across three model structures: BiLSTM, BiGRU, and CodeBERT. The results represent the average value of three independent runs, with minimal fluctuation (e.g., the standard deviation of AUC is approximately $10^{-5}$). We first use comprehensive metrics, including F1-score and AUC, to analyze the MIA performance results, revealing consistent trends across both MLP and CNN adversaries. 

In \textit{Feature 5}, \textit{Feature 6}, and \textit{Feature 8} consistently achieve the best attack performance, with AUC values often exceeding 0.99 across all target models. For instance, using MLP, \textit{Feature 6} reaches an AUC of 0.9996 on BiLSTM and 0.9999 on BiGRU, while \textit{Feature 8} achieves 0.9990 on BiGRU and 0.9927 on CodeBERT. These findings suggest that the combination of logits and loss is highly indicative of membership, and adding either confidence scores or embedding vectors further enhances the attack signal. CNN-based adversaries exhibit similar performance, indicating that these combinations are robust across model architectures. This may seem inconsistent with RQ2, where loss and logits show only marginal distributional differences between members and non-members. However, even weak per-feature signals can interact non-linearly when combined, revealing patterns of overfitting that attack models can exploit. Both CNNs and MLPs can capture these interactions, enabling strong attack performance despite the subtlety of individual signals.

Besides, \textit{Feature 2} alone already provides a relatively strong signal, with MLP-based AUCs of 0.6342, 0.4688, and 0.7259 on BiLSTM, BiGRU, and CodeBERT, respectively. While not as effective as the multi-feature combinations, this indicates that loss values alone encode a non-negligible amount of membership information. \textit{Feature 4}, which combines loss and confidence, offers no substantial improvement over using loss alone, highlighting the limited additional value of confidence in the absence of logits. Additionally, \textit{Feature 1}, \textit{Feature 3}, and \textit{Feature 7} perform close to random guessing, with AUCs around 0.5 regardless of the attack or target model architecture.

Beyond AUC, we also observe variations in precision and recall across feature sets. Some feature combinations (e.g., \textit{Feature 1} on BiLSTM and CodeBERT) yield extreme values, such as 0 or 1, which reflect that the attack model predicts all samples as either members or non-members. This commonly occurs when the input features lack sufficient discriminative power, leading adversaries to collapse to trivial decision rules.

Overall, the attack results corroborate the insights derived from prior sections: loss is a key signal for MIA, and its combination with logits, either with or without embeddings, amplifies the leakage. These findings emphasize the importance of carefully managing model outputs for security.

\subsection{Visualization of Membership Signals}
To gain deeper insight into the factors that lead to the success of MIA, we visualize the feature distributions of member and non-member samples on the target testing set, using t-distributed Stochastic Neighbor Embedding (t-SNE), a non-linear dimensionality reduction technique that preserves local structure while mapping high-dimensional data into a low-dimensional (typically 2D) space \cite{maaten2008visualizing}. This allows us to intuitively inspect the separability between member and non-member samples in the attack feature space. We compare three feature sets: \textit{Feature 2}, \textit{Feature 6}, and \textit{Feature 8} in Figure \ref{fig:tsne-visualization}, which progressively include more output-level signals from the target model, enabling a clear comparison of their separability. To quantify the visual separation between members and non-members, we compute the centroid distance for each plot, defined as the Euclidean distance between the mean feature vectors of the two classes. A larger centroid distance indicates greater inter-class separation, which correlates with stronger MIA performance.

In Figure \ref{fig:tsne_loss}, where only the loss is used (\textit{Feature 2}), the separation between member and non-member samples is weak across all target models. The centroid distances, 30.046 (BiLSTM), 33.022 (BiGRU), and 36.593 (CodeBERT), remain relatively small, and significant overlaps are observed in all three subplots. Correspondingly, the attack model performance based on \textit{Feature 2} is limited, with AUC scores around 0.63 to 0.78 across BiLSTM and CodeBERT models, but remains lower than 0.5 in BiGRU (see Table \ref{tab:mlp_mia} and \ref{tab:cnn_mia}). These results suggest that although loss carries some membership signal, its effectiveness is model-dependent. In BiGRU, the weaker separation may stem from its smoother training dynamics, which lead to similar loss values for both member and non-member samples. Without sharp overfitting behaviors or high-loss outliers, the loss alone fails to provide strong discriminative signals, resulting in poor attack performance.

\textit{Feature 5} and \textit{Feature 6} yield similarly strong attack performance and both significantly improve member–non-member separation, confirming that logits and loss are the dominant signals. Since the addition of confidence in \textit{Feature 6} provides only marginal gains over \textit{Feature 5}, their t-SNE visualizations show no substantial differences. Due to space constraints, we only visualize \textit{Feature 6} in Figure~\ref{fig:tsne_conf}. Although confidence offers limited additional benefit, its inclusion slightly enhances attack robustness when combined with logits and loss. These comparable distances align with their similarly high AUC scores under both MLP and CNN attacks.

\textit{Feature 8} further enhances separability in the CodeBERT model by incorporating embeddings alongside output-layer signals. The visualizations in Figure \ref{fig:tsne_emb} demonstrate that member and non-member samples form almost entirely disjoint clusters in the CodeBERT. The centroid distances (67.580) for \textit{Feature 8} are the highest among the three, and the attack AUCs remain competitive. These findings suggest that embeddings exhibit strong membership patterns, particularly when combined with final-layer outputs.

Overall, the visualization and performance results are highly consistent, where features that produce better class separation in the latent space also lead to more successful MIA. This alignment provides an intuitive understanding of why attacks succeed and highlights the underlying risk posed by certain types of model outputs and internal representations.

\begin{tcolorbox}[colback=gray!30, colframe=gray!70]
    \textbf{\textit{Answer to RQ3}}: Features combining logits, loss, and optionally embeddings or confidence consistently achieve superior attack performance.
\end{tcolorbox}

\section{Defense} \label{defense}
Motivated by the findings in RQ3, where logits and loss consistently emerge as the most privacy-leaking output for MIA, we propose a lightweight defense mechanism named Noise-based Membership Inference Defense (NMID). The core idea is to suppress potential membership signals by perturbing these vulnerable outputs, including logits and loss, before exposure, thereby mitigating the membership inference risk in both black-box and gray-box settings.

\subsection{Defense Design}

NMID is built upon the following two principles:

\textbf{(1) Output masking.}
We introduce a configurable mechanism to modify the final logits and loss values before they are returned or logged, aiming to suppress membership-related signals. Specifically, we implement a hard masking strategy that directly alters the most informative parts of these outputs. For logits, we identify the maximum logit value for each sample (i.e., the predicted class) and overwrite it with zero using index-wise masking: 
\[
\mathbf{z}_i[\arg\max_j \mathbf{z}_i[j]] \leftarrow 0 \quad \forall i,
\]
where $\mathbf{z}_i$ is the logit vector for sample $i$. This effectively suppresses the dominant class score and reduces the model’s confidence footprint associated with member predictions. For loss values, extreme numerical outputs are clamped to suppress large gradients that may leak membership status. These modifications are lightweight and can be seamlessly integrated into most model architectures.

\begin{figure}[t]
\centering
\includegraphics[width=0.9\linewidth]{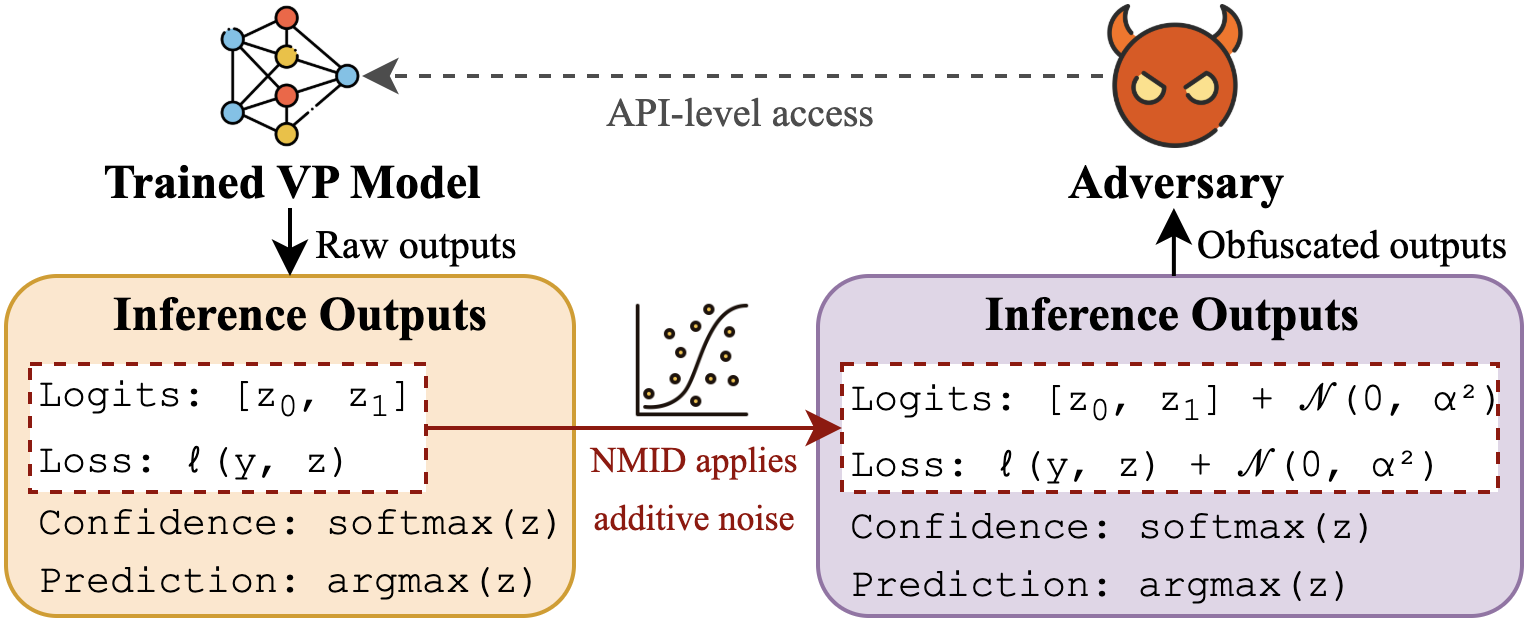}
\caption{NMID integrates noise injection modules into the logits and loss output.}
\label{fig:nmid}
\end{figure}

\textbf{(2) Additive Gaussian noise injection.}
To directly obfuscate attack-relevant signals, we inject zero-centered Gaussian noise into the outputs:
\begin{equation}
    \tilde{z}=z + \mathcal{N}(0, \alpha^2),
\end{equation}
where $z$ denotes the original logits or loss, and $\alpha$ is a tunable noise scale controlling the perturbation strength. This smoothing step generalizes the defense to resist stronger adversaries by introducing stochasticity into otherwise deterministic outputs. By varying $\alpha$, system administrators can flexibly control the trade-off between privacy and utility. This defense can be seamlessly integrated at the model’s output interface (e.g., prediction or evaluation module). The noise injection process is illustrated in Figure~\ref{fig:nmid}. In our experiments, we set $\alpha \in {3, 5, 10}$ to represent low, medium, and high noise intensity. These values are consistent with prior works \cite{li2021membership}, which demonstrated that Gaussian noise within this range achieves a balance between privacy and utility preservation.

Importantly, NMID is applied at the output interface during inference and does not modify the internal prediction logic of the model. The prediction labels are determined using the original logits before any masking or noise injection is performed. The perturbed outputs are only used for external exposure to ensure the model predictions remain consistent.

\begin{figure*}[t]
    \centering
    \begin{subfigure}[t]{0.47\textwidth}
        \centering
        \includegraphics[width=\textwidth]{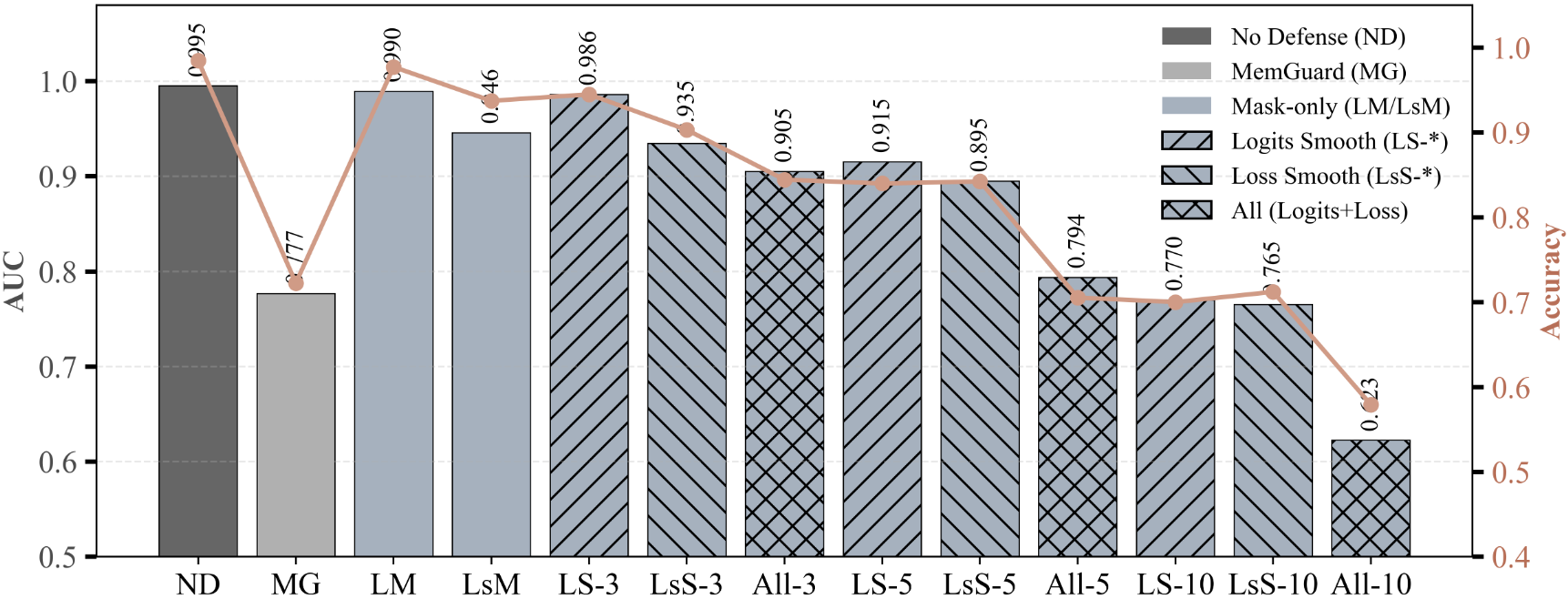}
        \caption{MLP Model.}
        \label{fig:eval_mlp}
    \end{subfigure}
    \begin{subfigure}[t]{0.47\textwidth}
        \centering
        \includegraphics[width=\textwidth]{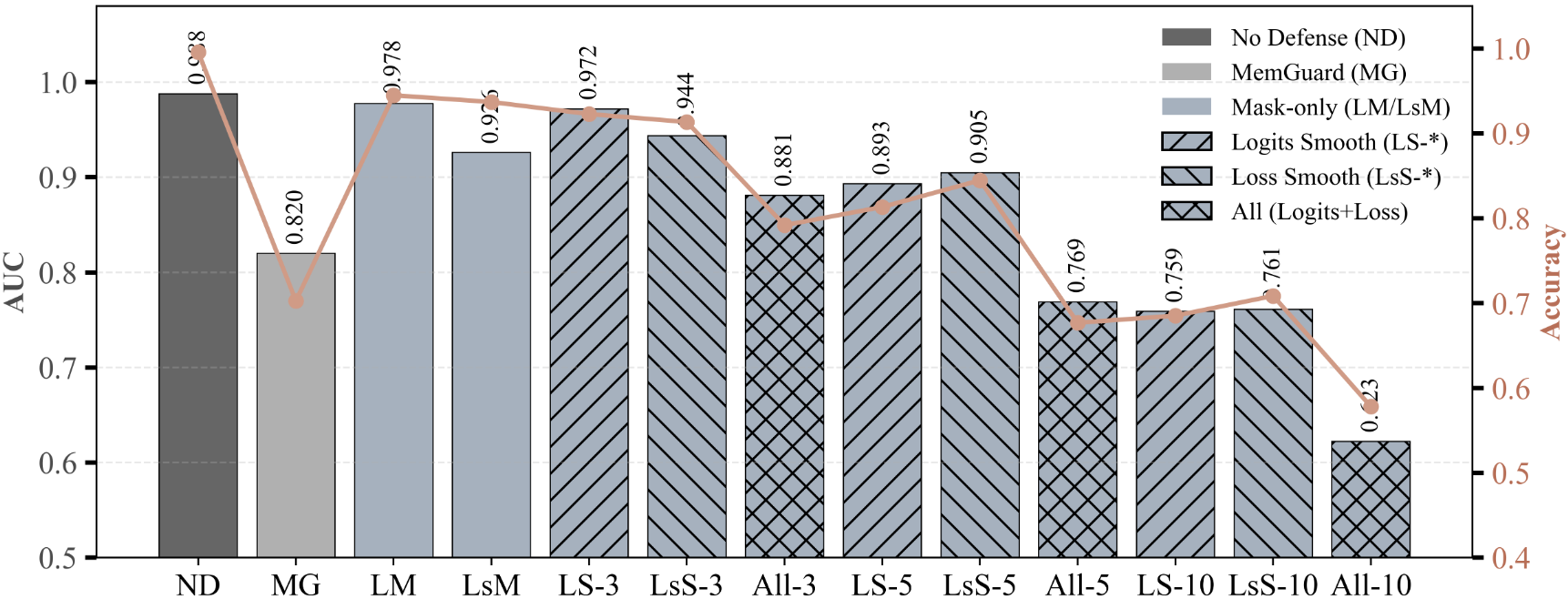}
        \caption{CNN Model.}
        \label{fig:eval_cnn}
    \end{subfigure}
    \caption{Attack Accuracy (Line) and AUC (Bar) under Defenses for different attack models.}
    \label{fig:eval_mlp_cnn}
\end{figure*}

\subsection{Deployment Flexibility}
NMID is a post-training defense that requires no retraining or architectural changes, making it suitable for real-world scenarios where access to model internals is restricted. It can be deployed as a lightweight wrapper at inference time, modifying logits and loss outputs through masking and noise injection. This allows seamless integration into existing machine learning pipelines, including edge devices and black-box APIs. Since NMID introduces perturbations only at inference, it preserves training efficiency and model performance. Moreover, it is robust to gray-box and black-box threat models, as it does not assume knowledge of the internal model structure.

\subsection{Baseline}
Common defenses against MIA include training-time strategies like DP-SGD \cite{abadi2016deep} and inference-time approaches such as MemGuard \cite{jia2019memguard}. In this study, we adopt MemGuard as our primary baseline due to its compatibility with NMID’s post-training deployment setting. MemGuard perturbs model outputs by generating adversarial noise designed to mislead MIA classifiers and has been widely adopted in prior work. Specifically, we follow the standard configuration and train a defense classifier, an MLP with three hidden layers (256, 128, and 64 units, ReLU activations), for 100 epochs using a learning rate of 0.001. The classifier distinguishes between member and non-member samples based on softmax outputs, and perturbed logits are then optimized to reduce the attack model’s confidence in member predictions.

Although DP-SGD provides formal privacy guarantees by injecting noise during training, it requires full model retraining with gradient clipping, which is often impractical in real-world scenarios involving large-scale or proprietary systems. Moreover, prior studies \cite{hu2022membership} have shown that DP-SGD can degrade utility in complex classification. In contrast, both MemGuard and NMID are lightweight, inference-time defenses that do not require architectural modifications or retraining, making them more feasible for practical deployment. 

\subsection{Evaluation}
To evaluate the robustness of our proposed defense NMID, we conduct experiments under various settings and visualize the results in Figure \ref{fig:eval_mlp_cnn}, including both MLP and CNN attack models. We evaluate NMID on the CodeBERT-based VP model using \textit{Feature 6} (logits+loss+confidence), which was previously shown to yield the highest MIA success rate across all feature combinations and model architectures. This scenario represents a strong adversarial setting, where the model output is rich and highly informative, making it a challenging but representative case for defense evaluation. In each figure, the x-axis includes 13 defense settings: no defense (ND), MemGuard (MG), logits mask (LM), loss mask (LsM), and smoothing with increasing Gaussian noise intensities ($\alpha = 3, 5, 10$). Specifically, LM denotes masking the maximum logit value, LsM clamps extreme loss values, LS-x refers to applying Gaussian smoothing (with scale $\alpha=x$) to the logits, LsS-x applies Gaussian smoothing to the loss, and All-x in NMID applies smoothing to both logits and loss simultaneously. The left y-axis reports AUC, and the right shows accuracy. Bars indicate AUC, lines indicate accuracy. The legend uses solid gray for ND/MemGuard and hatched blue for NMID variants.

As shown in both figures, MemGuard provides moderate protection against MIA attacks by lowering both attack AUC and accuracy compared to the no-defense setting. Specifically, it reduces AUC from 0.9953 to 0.7767 (decreases by 21.96\%) in MLP and from 0.9879 to 0.8202 (decreases by 16.98\%) in CNN, with corresponding accuracies of 0.7226 and 0.7026. These results demonstrate that MemGuard can effectively confuse the adversary in the absence of more substantial perturbations. NMID, in contrast, shows more flexible and tunable defense capabilities across a range of configurations. In particular, under the All-10 configuration, it achieves the most substantial reduction in attack performance, with AUC values dropping by 37.46\% (MLP) and 36.99\% (CNN), and accuracies decreasing by 41.17\% and 41.95\%, respectively, which approaches random-guessing performance. Notably, MemGuard's performance is comparable to several NMID variants, such as LS-10, LsS-10, and All-5, which exhibit similar AUCs and accuracies. However, NMID demonstrates clear superiority under stronger perturbation settings, such as All-10, where both AUC and accuracy are further reduced. This indicates that NMID removes more membership signal than MemGuard’s adversarial perturbation, without requiring extra adversary knowledge.

Within NMID, we further analyze the role of masking versus smoothing. The mask strategies (LM, LsM) slightly reduce AUC and accuracy. LsM lowers AUC from 0.9953 to 0.9460 (MLP) and from 0.9879 to 0.9262 (CNN). These results suggest that while masking is insufficient to mitigate MIAs, loss masking is comparatively more effective than logits masking. This may be attributed to the loss value being a single scalar with less redundancy, making it more sensitive to masking operations. 

For smoothing-based defenses (LS-*, LsS-*), the results confirm their superior effectiveness in disrupting attack performance. Even at a moderate noise scale of $\alpha=5$, attack metrics decline substantially, indicating that Gaussian smoothing introduces sufficient uncertainty to undermine adversarial decision boundaries. The strongest protection is achieved under the full NMID configuration (All-10), where both logits and loss outputs are perturbed. This setting consistently yields the lowest attack performance across all metrics (e.g., AUC = 0.6225, Accuracy = 0.5781), confirming that joint perturbation is essential for eliminating residual membership signals and outperforming partial smoothing or masking strategies.

We also evaluate the effect of increasing noise scale $\alpha \in \{3, 5, 10\}$. As $\alpha$ increases, both AUC and accuracy generally decline, indicating enhanced privacy at the cost of utility. This trend is exceptionally consistent in MLP-based attacks, where higher noise leads to reduced model confidence and stronger obfuscation. However, in the CNN-based attacks, we observe a few deviations from this trend. Specifically, the LsS-5 configuration shows slightly higher AUC and accuracy than LsS-3. These exceptions suggest that the effect of noise is not strictly monotonic across all configurations and may interact non-linearly with specific smoothing strategies and attack architectures. Nonetheless, the overall pattern validates the necessity of applying combined perturbations.

\section{Discussion} \label{sec:discussion}
\subsection{Why NMID Mitigates Membership Leakage?}
MIAs succeed because modern neural networks tend to memorize training data and behave differently on seen versus unseen inputs. This overfitting often manifests in sharper logits, lower loss, and higher confidence for member samples \cite{shokri2017membership}. These differences, though subtle per instance, become statistically significant across large samples, enabling adversaries to train classifiers that reliably infer membership. Moreover, the determinism of model outputs exacerbates this risk, especially when evaluation loss is logged during deployment, potentially leaking sensitive training signals.

Our proposed NMID defense addresses these vulnerabilities through two complementary mechanisms. Masking flattens the structure of logits or suppresses the range of loss values, directly targeting the structural overconfidence learned by the model. However, masking alone is insufficient, as relative patterns and residual signals may persist. The smoothing via additive Gaussian noise injects controlled randomness into the output, disrupting the stability of membership-specific patterns. This stochasticity forces the adversary to generalize across noisy separable inputs, effectively lowering their predictive performance. Since NMID operates externally to the model’s decision process, it preserves VP performance while selectively obfuscating signals exploitable by adversaries.

While both MLP and CNN adversaries exhibit strong performance under the no-defense setting, our results show that CNN-based adversaries experience more severe performance degradation across all noise levels. This trend is consistent across intermediate noise scales as well, indicating that CNNs are generally more sensitive to perturbations applied to output signals. One possible explanation is that CNNs, while effective at capturing localized patterns in high-quality data, may overfit to specific statistical regularities that are disrupted by Gaussian noise, making them less adaptable under stochastic defenses. These findings suggest that defenses like NMID are particularly effective against convolutional adversaries, demonstrating the importance of evaluating multiple adversarial architectures when designing privacy-preserving mechanisms.

\subsection{Trade-offs Between Model Capacity and Privacy}
Our findings reveal practical implications for choosing and deploying VP models under privacy constraints. Beyond output modality, leakage patterns vary notably across model architectures. Transformer-based CodeBERT consistently exhibits higher attack success, especially when only loss is exposed (AUC$\approx$0.72), while BiGRU performs at a near-chance level (AUC$<$0.5) under the same condition. We attribute this to two factors: (1) CodeBERT’s large capacity and self-attention allow detailed memorization, amplifying membership signals; (2) recurrent models enforce stronger sequence-level regularization, smoothing per-sample loss, and weakening single-feature attacks. However, when richer feature combinations are available, all architectures become vulnerable, indicating that capacity mainly affects individual signal separability rather than joint leakage patterns. These results suggest that while large pre-trained models like CodeBERT improve vulnerability detection, they also pose higher privacy risks due to stronger memorization. We recommend combining them with lightweight output-level defenses such as NMID, which achieves strong protection with minimal performance loss, helping balance utility and privacy in practice.

\subsection{Data Source Diversity and Privacy Implications}
Finally, while our dataset design integrates both synthetic samples from SARD and real-world samples from NVD to increase diversity and scale, we acknowledge that the two sources may exhibit different statistical characteristics. For instance, SARD samples tend to follow more templated or repetitive patterns, potentially leading to more substantial memorization effects, whereas NVD samples may introduce greater structural variation and noise due to their real-world origin. Although our study does not explicitly isolate their contributions, the combined use helps simulate realistic training scenarios where data sources are heterogeneous. We consider further dissecting the impact of dataset provenance on membership leakage an important direction for future work.

\section{Threat to Validity}

\textbf{Internal Validity.}
A key threat is unintended data leakage between shadow and target models. We enforce strict isolation to ensure no sample overlap. While CWE types appear in both member and non-member subsets to preserve diversity, overall CWE distributions are kept similar across shadow and target to avoid structural drift. Member and non-member splits are balanced to reduce bias. Another concern is whether the shadow model faithfully approximates the target's decision boundary. We mitigate this by using identical architectures, training settings, and fixed random seeds, and by averaging results over multiple runs.

\textbf{External Validity.}
Our experiments use three representative VP models (BiGRU, BiLSTM, CodeBERT) and real-world datasets (SARD, NVD), supporting generalization within the software security domain. However, findings may not directly extend to tasks in vision or NLP. NMID targets gray- and black-box settings where outputs, such as logits or loss, are observable, which is typical in ML inference APIs. In more restrictive settings that expose only hard labels, further adaptations would be required.

\textbf{Construct Validity.}
Our attack assumes adversaries can access outputs such as logits, loss, and confidence, commonly exposed in realistic deployments. Prior work shows these features are informative for MIA \cite{hu2022membership}, which our experiments confirm. We evaluate MIA using two representative architectures (MLP and CNN) that cover different learning paradigms. While more complex attackers may exist, our selected models serve as standard MIA baselines \cite{shokri2017membership}. NMID's consistent defense performance across both adversaries supports its generality.
\section{Related Work}

\textbf{Membership Inference Attacks.}
MIAs exploit differences in model behavior between training (member) and non-training (non-member) samples to infer membership status. Since the seminal work by Shokri et al. \cite{shokri2017membership}, many attack strategies have been proposed, such as entropy-based MIA \cite{hu2022membership, tang2022mitigating}. Recent surveys offer comprehensive overviews of attack methodologies, coverage, and defensive techniques in gray-box and black-box settings \cite{niu2024survey, hu2023defenses}. Notably, MIAs have been extended to large-scale models such as LLMs, revealing privacy risks across diverse architectures \cite{mattern2023membership, amit2024sok}.

\textbf{Defenses against MIAs.}
Common defenses include differential privacy (DP) \cite{abadi2016deep, zanella2023bayesian}, adversarial perturbation \cite{jia2019memguard}, and output noise injection \cite{leemann2023gaussian}. While training-time methods like DP and label smoothing offer theoretical guarantees, they require model retraining and often degrade utility in classification tasks \cite{hu2022membership}. In contrast, inference-time defenses such as MemGuard and NMID can be deployed post hoc without architectural modification. NMID builds on this line by introducing lightweight masking and calibrated Gaussian noise on logits and loss, providing effective protection without requiring additional knowledge from the adversary.

\textbf{Vulnerability Prediction.}
Deep learning-based VP has matured considerably, with LSTM/GRU and transformer architectures (e.g., CodeBERT) being widely adopted \cite{kalouptsoglou2023software}. Hybrid approaches that integrate static analysis and neural networks have also shown promise in predicting the presence and severity of vulnerabilities \cite{shiri2024systematic}. In particular, transformer-based CodeBERT models consistently outperform RNN-based baselines on C/C++ code tasks, while bidirectional LSTM/GRU variants remain competitive in certain settings \cite{xia2024vulcobert}.

\textbf{Comparison with MIA on Code Completion Models.}
Recent studies have examined MIA in the code completion models. Yang et al. \cite{yang2024gotcha} proposed Gotcha, an MIA method for generative code models like CodeGPT, using generated output sequences and auxiliary matching signals to infer membership. Wan et al. \cite{wan2024does} introduced CodeMI, which trains shadow models to simulate black-box APIs and extracts ranked prediction vectors as MIA features. However, both works target generative settings where outputs are token sequences and rely on perplexity, token ranking, or sequence similarity to infer membership. In contrast, our work focuses on classification-based VP models, where outputs such as logits and loss are directly observable and carry strong membership signals. Moreover, existing works do not consider any defense mechanisms against such privacy risks. Our work not only demonstrates that VP models are significantly more vulnerable under standard output configurations but also proposes a deployable defense (NMID) that mitigates MIA risk with minimal utility loss. To our knowledge, this is the first study to bridge MIA risks and mitigation strategies in VP.

\section{Conclusion}

In this paper, we present the first comprehensive empirical investigation of MIAs targeting neural models for VP. Our results reveal that VP models, including BiGRU, BiLSTM, and CodeBERT, are highly susceptible to MIAs, especially when adversaries have access to intermediate outputs such as logits and loss. Under these conditions, attack models can achieve near-perfect inference performance, F1-scores and AUCs exceeding 0.95 in some configurations, indicating severe privacy risks inherent in widely used VP pipelines.

To address this challenge, we propose \textbf{NMID}, a lightweight defense framework that operates without retraining or architectural modification. NMID mitigates membership leakage by masking sensitive output patterns and injecting calibrated Gaussian noise into logits and loss vectors. Our extensive evaluations demonstrate that NMID consistently reduces attack success across diverse settings. When applied with a high noise scale ($\alpha=10$), the adversary’s F1-score drops from over 0.95 to as low as 0.56, approaching the level of random guessing. 

Our findings underscore the crucial need for privacy-aware deployment of VP models and demonstrate that principled yet straightforward defenses can offer robust protection against inference threats.

\bibliographystyle{IEEEtran}  
\bibliography{ref}

\end{document}